\documentclass{article}

\usepackage{arxiv}

\usepackage[utf8]{inputenc} 
\usepackage[T1]{fontenc}    
\usepackage{hyperref}       
\usepackage{url}            
\usepackage{booktabs}       
\usepackage{amsfonts}       
\usepackage{nicefrac}       
\usepackage{microtype}      
\usepackage{lipsum}

\usepackage[bottom]{footmisc}

\usepackage{setspace} 
\usepackage[utf8]{inputenc} 
\usepackage{amsmath, amsthm, amssymb}	
\usepackage{amssymb}               
\usepackage{graphicx}              
\usepackage{epstopdf}
\usepackage{commath}
\usepackage{lscape}
\usepackage{amsfonts}
\usepackage{icomma} 
\usepackage[small,bf,hang]{caption}

\usepackage{algorithm}
\usepackage[noend]{algpseudocode}

\usepackage{booktabs}
\usepackage{pgfplots}
\usepackage{pgfplotstable}
\usepackage{threeparttable}

\author{
	Yanou Ramon\\
	Dpt. of Engineering Management\\
	University of Antwerp\\
	\And
	Tom Vermeire\thanks{Yanou Ramon and Tom Vermeire contributed equally.}\\
	Dpt. of Engineering Management\\
	University of Antwerp\\
	\And
	Olivier Toubia\\
	Dpt. of Marketing\\
	Columbia University\\	
	\And
	David Martens \\
	Dpt. of Engineering Management\\
	University of Antwerp\\
	\And
	Theodoros Evgeniou \\
	Decision Sciences and Technology Management\\
	INSEAD Paris \\
}

\begin{document}
	
\title{Understanding Consumer Preferences for Explanations Generated by XAI Algorithms}
\maketitle

\keywords{Algorithmic Decision Making; Explainable Artificial Intelligence; Algorithmic Explainability; Choice-Based Conjoint Analysis}

\begin{abstract}
Explaining firm decisions made by algorithms in customer-facing applications is increasingly required by regulators and expected by customers. While the emerging field of Explainable Artificial Intelligence (XAI) has mainly focused on developing algorithms that generate such explanations, there has not yet been sufficient consideration of customers' preferences for various types and formats of explanations.
We discuss theoretically and study empirically people's preferences for explanations of algorithmic decisions. We focus on three main attributes that describe automatically-generated explanations from existing XAI algorithms (format, complexity, and specificity), and capture differences across contexts (online targeted advertising vs. loan applications) as well as heterogeneity in users' cognitive styles. Despite their popularity among academics, we find that counterfactual explanations are not popular among users, unless they follow a negative outcome (e.g., loan application was denied). We also find that users are willing to tolerate some complexity in explanations. Finally, our results suggest that preferences for specific (vs. more abstract) explanations are related to the level at which the decision is construed by the user, and to the deliberateness of the user's cognitive style. 
\end{abstract}

\section{Introduction}
\label{Section:Introduction}
Algorithms are widely used to automate firm decisions in a variety of business domains. Classic examples are credit scoring and loan decisions in the financial sector~\cite{CreditScoring}, data-driven recruitment (e.g., automated resume screening \cite{ResumeScreening}), targeted advertising and marketing-based recommendations (e.g., which products to buy \cite{JulieExplanations}, which restaurants to visit \cite{JulieExplanations}, or which movies to watch \cite{Netflix}). So far, the question of generating explanations to customers of why these algorithms make certain decisions has received scant attention in the Marketing literature. However, regulatory requirements and increasing customer expectations for responsible Artificial Intelligence (AI)~\cite{EIU2020,ProjectExplain2019} raise new research questions about how to best design explanations for algorithmic decisions. For example, the General Data Protection Regulation (GDPR)~\cite{GDPR22} 
notes the ``right to explanation'' when decisions, such as recommendations of products or displays of advertisements, are made by algorithms. The recent European Commission's Digital Services Act~\cite{DSA2021} further emphasises this by noting that recipients of online advertisements should have access to ``meaningful explanations of the logic used'' for ``determining that specific advertisement is to be displayed to them'' (paragraph 52). Beyond increasing regulatory pressures, there may be additional reasons for companies to make such explanations available. For example, people are often averse or skeptical to rely on algorithms \cite{Dietvorst2018,Yeomans2019}. Prior work indicates that people tend to be less averse to recommender systems when they are provided with explanations about how they work \cite{Yeomans2019,cadario2021understanding}, and that in advertising, people tend to accept personalized advertisements more easily when they understand why they are seeing an ad \cite{Summers}. Along the same lines, \cite{morey2015customer} make the call for companies to design products and services with transparency in mind.  

While explaining AI decisions is becoming increasingly important, we know little about what makes explanations more or less preferable to people. For example, should explanations report the factors that contributed to an algorithmic decision (feature importance), or factors that would have led to a different decision (counterfactual)? 
How complex should explanations be? Should explanations mention specific, concrete pieces of information (e.g., ``you are seeing this ad because you visited this particular website'') or more abstract reasons (e.g, ``you are seeing this ad because you are generally interested in sports'')? And of course, how do the answers to these questions vary based on the context, the type of algorithmic decision being explained, and the characteristics of the individual to whom the explanation is given?

The field of Explainable AI (XAI) has emerged as a subfield at the intersection of Information Systems and Computer Science, with no representation (to the best of our knowledge) in the Marketing literature. XAI has mostly focused on developing algorithms that automatically generate explanations (see \cite{XAIreview} and \cite
{2018Guidotti} for overviews of explanation methods proposed in the literature). Researchers typically make explicit or implicit assumptions about the type of explanations that end users would like to see. 
However, in order for XAI to gain wide adoption, increase understanding, trust and acceptance of technologies~\cite{Yeomans2019}, and have a positive impact on consumers, it is essential to test these assumptions empirically and to gain a deeper understanding of people's preferences for explanations of algorithmic decisions \cite{Doshi2017,MillerInmates,Miller2019,lu2020good,Keane2021,Riveiro2021}. We argue that the Marketing literature has the potential to make important contributions on this front, both empirically and theoretically. Empirically, the preference measurement methods that have been vetted in Marketing over the past decades are well poised to provide precise, quantitative insights on users' preferences for explanations. Theoretically, Marketing can contribute its fundamental understanding of consumer behavior and consumer preferences in order to develop a framework for thinking about key factors that drive preferences for explanations, and reach general conclusions on what type of explanations would be preferred by different types of users in different contexts. 

In this article, we establish a bridge between the field of Marketing and the emerging field of XAI, by studying people's preferences for key characteristics of explanations generated by today's XAI algorithms. Our theoretical discussion leads to a series of predictions which we test empirically in two studies that both rely on Choice-Based Conjoint Analysis (CBC)~\cite{Orme2000,Allenby2005}. We stay close to current XAI algorithms and focus on three main attributes that characterize explanations generated by these algorithms: explanation format (counterfactual vs. feature importance), explanation complexity (low vs. moderate), and feature specificity (low- vs. high-level features). Prior research has shown that context plays an important role in the perception and appreciation of explanations~\cite{ProjectExplain2019}. Therefore, we study how preferences for explanations vary across contexts, by studying and contrasting explanations related to targeted advertising vs. loan applications. 

We find that counterfactual explanations, despite being often favored by XAI researchers \cite{Martens2014,Wachter2017,Miller2019,Ramon2020}
, are in general less popular. Consistent with our predictions, we find that aversion to counterfactual explanations is reduced when these explanations are provided following a negative decision (e.g., loan application was denied). In addition and contrary to common wisdom, we find that users are able to tolerate a moderate level of complexity. Finally, we find that preferences for specificity are context-dependent, in a way that is consistent with construal level theory \cite{trope2007construal,lambrecht2013does}. Users prefer specific features in the context of loan application (which happens late in the purchase process, after the consumer has already applied for a loan, and therefore should be construed at a lower level by the consumer) and less specific features in the context of targeted advertising (which is relevant earlier in the purchase process, and therefore should be construed at a higher level by consumers). We find that users who have a more deliberative cognitive style (higher score on the cognitive reflection test) have a stronger preference for lower-level, specific features. 
Beyond their intrinsic value, our findings emphasize the importance of empirically testing assumptions on user preferences before embarking on the development and deployment of XAI tools. Based on our findings, it appears the XAI literature has converged to formats of explanations that may not be well aligned with users' preferences, at least in some contexts. 

The article is organized as follows.  We first review some relevant literature from Marketing and XAI, introduce the key characteristics of XAI explanations that we study, and discuss how they relate to our knowledge of consumer behavior and consumer preferences. We then discuss our experimental setup, followed by the results of our two studies. We close with a discussion about future research directions in Marketing related to XAI.

\section{Relevant Literature}
\vspace*{5mm}
\subsection{Privacy and Transparency in Online Advertising}
In the context of online advertising, some relevant research in Marketing has studied how the use of personal information in targeting relates to issues of transparency and privacy  (e.g., )\cite{goldfarb2011online,goldfarb2011privacy,goldfarb2012shifts,kim2019seeing, 2021JohnsonShriver
}).
While this literature generally focuses on \textit{which} information a firm relied upon when targeting customers, we focus on the complementary question of \textit{how} this information should be presented to users in explanations. 
In particular, we investigate people's preferences for different types of explanations that have been proposed in the XAI literature. Compared to this literature, our study also extends beyond online advertising. 

A paper particularly relevant to ours is \cite{lambrecht2013does}, who show that consumers respond more favorably to higher-level (resp., lower-level) product information when an online targeted ad is shown early (resp., late) in the purchase process, due to the construal level of consumer's preferences getting more narrow as the consumer progresses through the purchase process. We rely on this study to predict that consumers prefer explanations based on features that are less (vs. more) specific for algorithmic decisions that happen early in the customer's purchase process (e.g., online advertising) vs. later (e.g., loan decisions). 

\subsection{Human Attitudes towards AI Decisions}
Another relevant literature has studied human acceptance of AI. \cite{dietvorst2015algorithm} coined the term \textit{algorithm aversion}, to describe humans' general preference for human forecasters over algorithms. \cite{Dietvorst2018} show that algorithm aversion can be reduced by giving people some control over the algorithm's forecast. \cite{longoni2019resistance} show that consumers are resistant to AI in healthcare when they perceive that AI fails to account for each patient's unique characteristics. \cite{longoniword} show that consumers are more welcoming of (resistant to) recommendations for utilitarian (hedonic) products. Along the same lines, \cite{Yeomans2019} find that people are more averse to relying on AI systems than on human judgments, and that this aversion partly stems from the fact that people perceive the human recommendation process as easier to understand. In contrast to this literature, we do \textit{not} compare 
explanations coming from AI vs. humans, but rather study, within the realm of AI, how consumers perceive automatically-extracted explanations provided about AI decisions, i.e., we compare AI explanations to other AI explanations. 

\subsection{Explainable AI}

In the current XAI literature, consumers' preferences for explanations are rarely considered \cite{MillerInmates,Riveiro2021}, and the majority of methods proposed in the XAI domain have arisen from the point of view of computer/data scientists. This relates to the phenomenon of the ``inmates running the asylum'' \cite{MillerInmates}: AI researchers are building explanatory agents for themselves, rather than for the intended users. \cite{MillerInmates,Miller2019} argues that explainable AI is more likely to succeed if researchers and practitioners understand, adopt, implement, and improve models from the vast and valuable bodies of research in philosophy, psychology, and cognitive science, and if the evaluation of these models is focused more on people than on technology. Along similar lines, \cite{Keane2021} argue that the neglect or poor conduct of user studies is an important shortcoming in existing XAI research.\footnote{Prior XAI work with user studies either evaluated a specific explanation method against a no-explanation control or measured the (perceived) usefulness and comprehensibility of explanations (see \cite{Keane2021} for an overview).} 

Two studies in the XAI literature that are particularly relevant to our paper are  \cite{lu2020good} and \cite{Riveiro2021}. 
\cite{lu2020good} compare various formats of explanations using a survey (e.g., no explanation, verbal explanation, decision tree, neural network). Our approach is related, but we focus on a different set of attributes of explanations, which we vary simultaneously using a CBC experiment. 
\cite{Riveiro2021} empirically investigate respondents' preferences for factual and counterfactual explanations in the context of AI-assisted decision making (involving a text classification task) as a function of users' expectations of the AI predictions. They find that factual explanations are appropriate when expectations and output match, but that neither factual nor counterfactual explanations appear appropriate when expectations and output do not match. 
We also compare factual to counterfactual explanations, but we vary the context along other dimensions (positive vs. negative decisions, low vs. high stakes, online advertising vs. loan applications). 

\section{Theoretical Development}
We now turn to introducing the three key attributes of XAI explanations that we study in this paper, and discussing how each of them relates to the broader Marketing literature on consumer behavior and preferences. 


\subsection{Format}
\label{section:format}
\begin{figure}[h!]
	\centering
	\scalebox{0.3}{\includegraphics{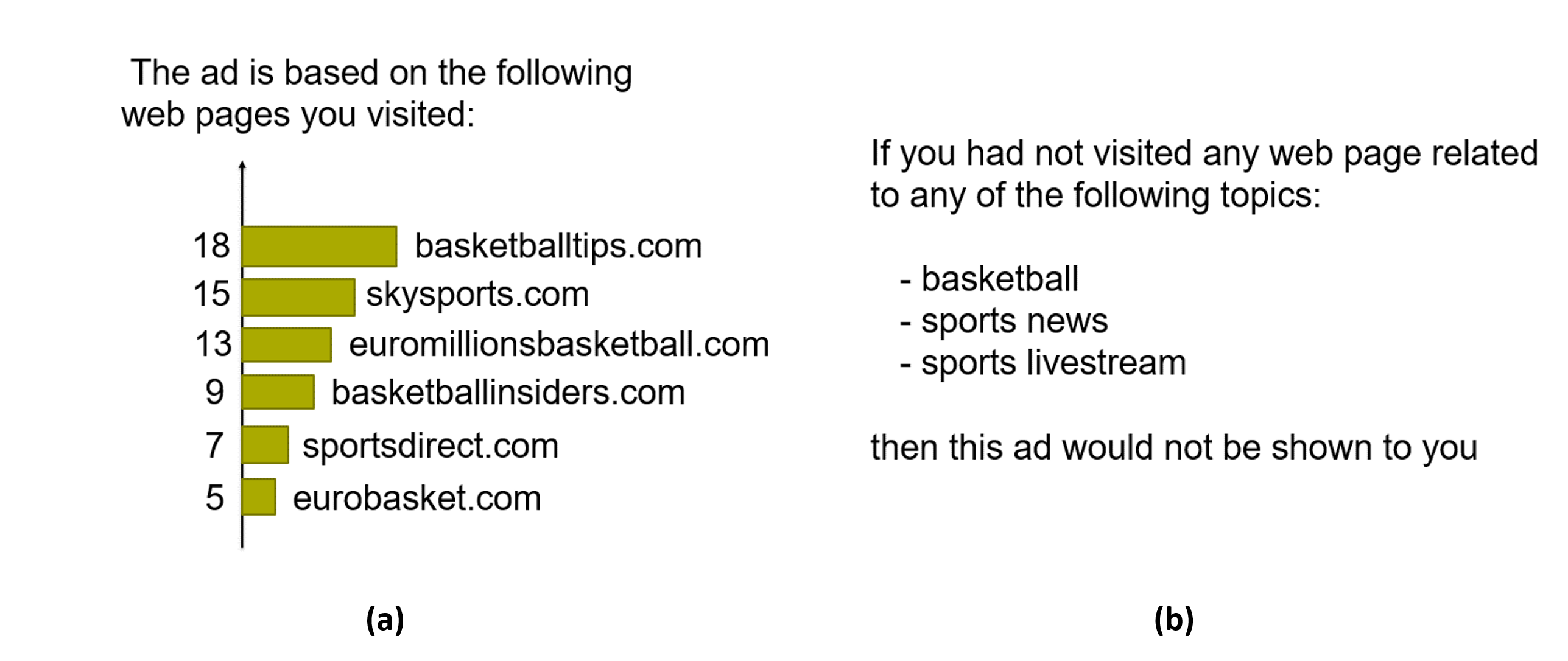}}
	\caption{Examples of explanations: a one-sided feature importance ranking with moderate complexity and low-level features (a) and a counterfactual explanation with low complexity and high-level features (b).}
	\label{fig:examples}
\end{figure}

Today's XAI methods come in a few different formats: explanations can be merely visual \cite{Apley}, rule-based \cite{CreditScoring,Martens2014,Ramon2020} or point to the most important features that led to a decision \cite{Ribeiro2016,Lundberg2017}.
A popular and widely-used method is known as feature importance ranking, wherein the model features are ranked according to their contribution to an algorithmic decision \cite{Ribeiro2016, Lundberg2017}.\footnote{Technically, features are ranked according to their contribution to a prediction score that led to the decision~\cite{fernandez2020}.} A distinction is made between rankings with only features that positively contributed to the decision and rankings that also contain features that had a negative contribution. In the remainder of this paper, we respectively call these one-sided and two-sided importance rankings (one-sided IR vs. two-sided IR). \textbf{Fig. \ref{fig:examples}(a)} shows an example of a one-sided feature importance ranking in the context of online targeted advertising.

Another explanation format that is gaining popularity is based on the notion of counterfactual explanations~\cite{Martens2014,Wachter2017,Karimi2020,Ramon2020}. Counterfactuals 
were first introduced in the XAI field by \cite{Martens2014} as a minimal set of feature changes that alter an algorithmic decision. \textbf{Fig. \ref{fig:examples}(b)} shows an example. It is important to note that the format of the explanation need not be linked to the actual algorithm that led to the decision. For example, a decision could be reached by a random forest model, and explained using a counterfactual. 

\begin{center}
	INSERT FIGURE \ref{fig:examples} HERE
\end{center}

While the counterfactual format of explanations is growing in popularity among XAI scholars \cite{Martens2014,Wachter2017,Miller2019}, with some arguing it closely aligns with recent regulatory initiatives like GDPR~\cite{Wachter2017}, the Marketing and Psychology literature suggest that it is unlikely to appeal to many consumers in many contexts. Indeed, according to the ``positive test strategy'' proposed by \cite{klayman1987confirmation}, humans have a natural tendency to favor and seek instances in which a certain property is expected to occur, rather than focusing on instances in which it does not occur. This fundamental property of human cognition is linked to the well known confirmation bias \cite{wason1960failure,nickerson1998confirmation}. \cite{skov1986information} and \cite{devine1990diagnostic} show that people tend to prefer questions that would yield a positive answer if a hypothesis is correct over questions that would yield a negative answer. \cite{evans1989bias} refers to ``a form of selective processing which is very fundamental indeed in cognition - a bias to think about positive rather than negative information'' (p. 42). This idea may be traced back to English Renaissance philosopher Francis Bacon, who noted in 1620 that ``it is the peculiar and perpetual error of the human understanding to be more moved and excited by affirmatives than negatives; whereas it ought properly to hold itself indifferently disposed towards both alike'' \cite[][p.36]{burtt1939english}. Due to this positivity bias, we expect counterfactual explanations, which explain to consumers under what conditions the decision would \textit{not} have been reached, to be in general less preferred than positive explanations that provide factors that contributed to the decision being reached. 


There are, however, situations in which we might expect counterfactual explanations to be relatively more popular (or less unpopular). In particular, consider cases in which the explanation is provided for a \textit{negative} decision, e.g., a loan application was denied. There are at least two reasons to predict that counterfactual explanations would be more desirable in such cases. First, in these cases a counterfactual explanation provides conditions under which a \textit{positive} decision would have been reached. Hence, positivity bias actually works in favor of counterfactuals in the case of negative decisions.  
Second, according to regulatory focus theory, a negative decision is likely to induce prevention focus and motivate users to avoid negative outcomes \cite{higgins1997beyond}. Research has shown that messages are more effective when they display regulatory fit, i.e., the regulatory focus of the message matches the regulatory focus of the message's recipient \cite{lee2004bringing,zhang2010does}. Accordingly, we should expect counterfactual explanations for negative outcomes, which explain how negative outcomes could have been avoided and are therefore consistent with prevention focus, to be relatively preferred, compared to counterfactual explanations for positive outcomes. 


\subsection{Complexity}
The complexity of a prediction model can be increased, for example, by adding more features in a linear model or by adding more paths to a decision tree \cite{Freitas}. A general assumption in the literature is that linear models with few parameters or rule-based models with few rules are more comprehensible (and less complex) than linear models with many parameters or rule sets with many different rules~\cite{Huysmans2011,Freitas}. We can extend this to the XAI literature and claim that explanations that contain more features are less comprehensible (and more complex) compared to concise explanations. Accordingly, we use the number of features in an explanation as a measure of the complexity of the explanation. We acknowledge that this is not the only measure of complexity, but it is a convenient proxy. 

How the complexity of an explanation might impact a user's subjective preference and appreciation for the explanation is an empirical question. Research in psychology points at the limited processing capacity of humans. \cite{Miller1956} states that, on average, people are capable of storing about seven pieces of information in their short-term memory. One may conjecture that the complexity of explanations should be subject to this same limitation, and we should expect preferences for explanations to go down as the number of features grows large~\cite{Sweller1988}. Moreover, research shows that people 
find simple explanations to be both better and more likely to be true \cite{Lombrozo2007,MillerInmates}. 

The more interesting question, on which we focus, is whether preferences are monotonically decreasing early on, or whether there exists a range over which preferences are actually increasing, i.e., consumers prefer moderate complexity to low complexity. Accordingly, in our experiments we study explanations with three and six features. For example, in the targeted advertising context, we experiment with explanations that contain three vs. six visited websites to explain why someone was targeted with an ad. The examples shown in \textbf{Fig. \ref{fig:examples}} have a respective number of features of six for the feature importance ranking and three for the counterfactual explanation. 

\subsection{Specificity}
\label{section:specificity}
The features that are used by the prediction model are the building blocks of the explanations. However, the original features used by the model might not always lead to the most comprehensible explanations. Some researchers have explored the idea of mapping the original features to a higher-level representation when explaining model decisions. In other words, they use another set of features (i.e., other than the original features on which the model makes decisions) to explain the model's decision making. For example, higher-level ``metafeatures'' for behavioral and textual data \cite{Ramon2020} (e.g., categories of websites like ``Marketing'' and ``Data Science'' instead of the websites themselves) can be used to explain decisions of a targeted advertising engine. 

In our experiments, we study explanations that contain low-level vs. high-level features. The counterfactual explanation in \textbf{Fig. \ref{fig:examples}} has high-level features (website categories), while the importance ranking has low-level features (individual websites).

\cite{lambrecht2013does} study specificity in the online advertising context of retargeting. In doing so, they invoke Construal Level Theory \cite{trope2007construal}, according to which higher levels of construal tend to represent information that is abstract and general, while lower levels of construal tend to represent information that is concrete and specific. \cite{lee2010value} show that messages that present information at abstract, higher levels are received more favorably by consumers who construe information at a higher level, and vice versa. That is, fit between the level at which an individual construes information and the level at which a message construes information, leads to more successful messages. Based on this literature, \cite{lambrecht2013does} hypothesize and find that ads shown early in the purchase process, when consumers' preferences are construed at an abstract, high level, are more effective when they present abstract information, and that ads shown later in the process, when consumers have more concrete preferences, are more effective when they present more specific information. Based on this finding, we can predict that consumers will have a preference for more specific, concrete features in contexts that reflect a lower level of construal. Such contexts include explanations for algorithmic decisions that take place late in the purchase process, e.g., whether a loan application was approved or denied (after the consumer already took the action of applying for the loan). In contrast, consumers should prefer more general, abstract features in contexts that reflect a higher level of construal. Such contexts include explanation for algorithmic decisions that take place early in a purchase process, e.g., online display advertising that typically occurs before a consumer has committed to making a purchase. 

\subsection{Application Context}
A recent study on understanding the expectations of individuals when explaining decisions of AI systems, conducted by the Information Commissioner's Office and Alan Turing Institute~\cite{ProjectExplain2019}, showed that context is key when explaining AI decisions to individuals. The reasons for and the importance of providing explanations can dramatically vary depending on what the decision is about. 


Given this, and given the literature reviewed in subsection \ref{section:specificity} that suggests that preferences for specificity should vary across contexts, we chose to include two contexts in our experiments: online advertising and loan applications. 
Most people have (significant) experience with both contexts, and hence are likely to be familiar with them. In addition, loan applications may be accepted or denied, hence the loan application context allows testing our prediction that counterfactual explanations should be relatively more popular following negative decisions. In addition, one of these two contexts, online advertising, is relevant in the early stages of a consumer's purchase process, while the other, loan application, is relevant after the consumer has already completed the purchase (i.e., application) process. Hence, these two contexts also lend themselves well to illustrating how context relates to preferences for specificity, along the lines predicted by construal level theory. Of course, many factors differ between loan applications and online advertising, which may also explain  differences in preferences between these two contexts. While we address some of these alternative explanations in Study 2, we can only claim that the differences we observe relative to preference for specificity between online advertising and loan applications are \textit{consistent} with the predictions from construal level theory, but we cannot claim that this is the only possible explanation. 

\subsection{Users' Cognitive Styles}
It has been shown that firms may improve the profitability of their online activities by personalizing content on the basis of users' cognitive styles \cite{hauser2009website}. This brings up the question of whether and how preferences for XAI explanations depend on users' cognitive styles. 

\cite{hauser2009website} argue that impulsive visitors to a website might prefer less detailed information, while deliberative visitors might prefer more information. Consistent with this, \cite{urban2014morphing} find that banner ads with detailed lists of product features are more effective with users who are more deliberative. In our context, this suggest that deliberative visitors might have a stronger preference for moderate complexity (i.e., more features) and for low-level (i.e., more concrete, detailed) features. 

\cite{hauser2009website} further propose that the cognitive reflection test (CRT) \cite{frederick2005cognitive} offers a valid measure of impulsivity vs. deliberativeness. CRT is a simple, very popular, three-item scale introduced by \cite{frederick2005cognitive}, shown to correlate positively with other measures of cognitive ability (e.g., IQ, SAT scores), while being a better predictor of decision making. \cite{frederick2005cognitive} argues that CRT specifically measures cognitive reflection, ``the ability or disposition to resist reporting the response that first comes to mind'' (page 35).\footnote{For example, if a ball and a bat together cost \$1.10 and the bat costs a dollar more than the bat, the first response that comes to mind may be that the ball costs 10 cents, when the correct response is 5 cents.} In our studies, we administer the CRT scale to all our respondents, to study how CRT correlates with preferences for explanations (see the survey details in the Web Appendix). 
While following \cite{hauser2009website} and \cite{urban2014morphing} in our use of CRT, we acknowledge that this scale has recently been criticized for being too closely related to numeracy~\cite{peters2020innumeracy}, and for being known to some online panelists~\cite{thomson2016investigating}. We leave the study of additional measures of cognitive styles to future research. 

\section{Research Methods}
\vspace*{5mm}
\subsection{Experimental Setup}
To measure people's preferences for explanations of algorithmic decisions, we conducted two studies that relied on CBC~\cite{Allenby2005}. Respondents were presented with a series of binary choice sets, where ``products'' were explanations for algorithmic decisions, and  ``attributes'' were the three characteristics of explanations reviewed in the previous section. Format had three levels: counterfactual explanation, feature importance ranking with one-sided information, and feature importance ranking with two-sided information. Complexity  had two levels: low complexity (three features) vs. moderate complexity (six features). Feature specificity had two levels: low-level vs. high-level features. 

Each choice question showed two explanations for an AI decision: one counterfactual explanation, and one explanation that was either a one- or a two-sided importance ranking. One of the explanations had low complexity, and the other had moderate complexity. One of the explanations had high-level features, and the other had low-level features. To maximize the information gathered, we varied these features according to a statistically efficient design for both studies, obtained from the leading conjoint software \textit{Sawtooth} (https://sawtoothsoftware.com/).  The set of CBC questions was \textit{not} adapted based on the respondents' profiles. That is, in a given condition, all respondents saw the same set of questions.

\begin{figure}[h!]
	\centering
	\scalebox{0.8}{\includegraphics{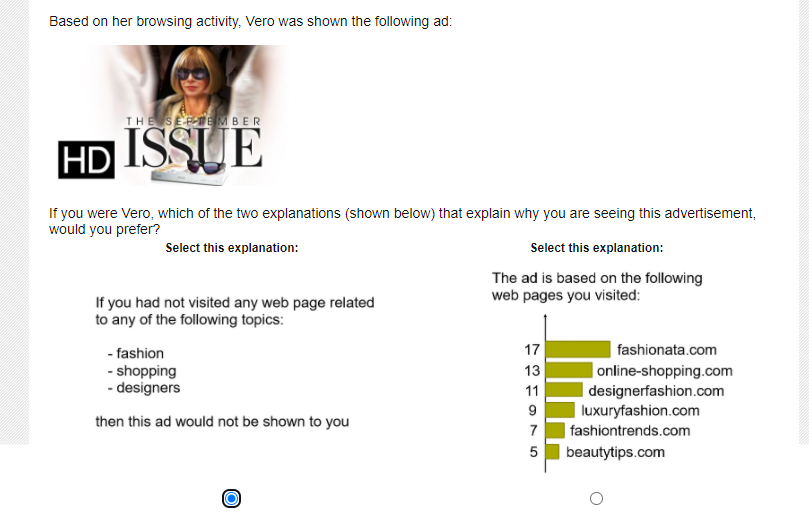}}
	\caption{Screenshot of a CBC question in the context of targeted advertising (Study 1).  A low-complexity counterfactual with high-level features (i.e., categories of websites) is shown on the left. A moderate-complexity, one-sided importance ranking with low-level features (i.e., websites) is shown on the right.}
	\label{fig:QuestionTA}
\end{figure}

\begin{figure}[h!]
	\centering
	\scalebox{0.8}{\includegraphics{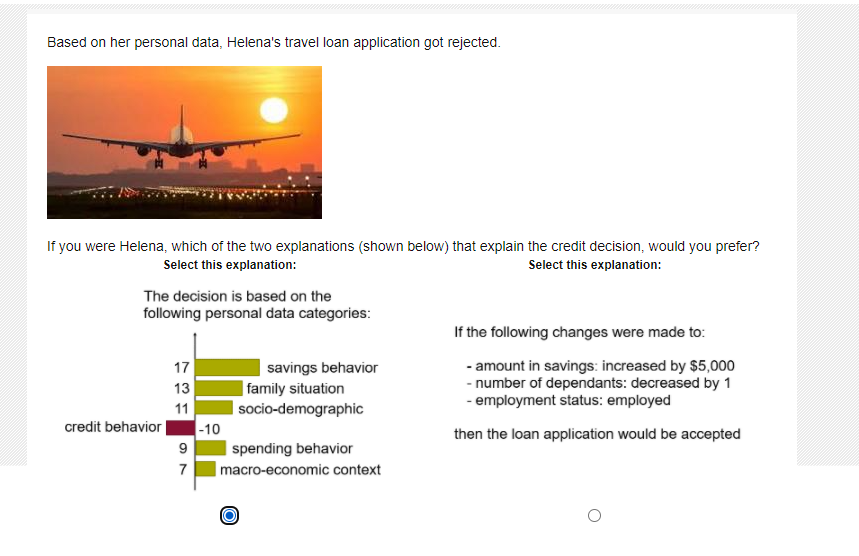}}
	\caption{Screenshot of a CBC in the context of loan application (Study 2, low-stake/negative loan decision). A high-complexity, two-sided importance ranking with high-level features is shown on the left. A low-complexity counterfactual with low-level features is shown on the right.}
	\label{fig:QuestionCS}
\end{figure}

Unlike a typical CBC study, our study required developing a context for each question (i.e, we could not simply tell respondents that one explanation had high-level features, we had to show them an actual explanation). Accordingly, we created one vignette for each choice question (e.g., ``Based on her browsing activity, Vero was shown the following ad:''). Examples of choice questions for targeted advertising and loan applications are shown respectively in \textbf{Figs. \ref{fig:QuestionTA}} and \textbf{\ref{fig:QuestionCS}}, and the full questionnaires are included in the Web Appendix. The choice of the vignette informed the specific explanations shown in the question (e.g., topics related to fashion, shopping and designers were mentioned in explanations for a fashion ad). For each vignette, we developed a relevant set of low-level features and a relevant set of high-level features, with a one-to-one mapping between the two sets. For example, in the vignette shown in \textbf{Fig. \ref{fig:QuestionTA}}, the three high-level features were developed to match the first three low-level features (e.g., ``fashionata.com'' is a low-level feature matched to the higher-level feature ``fashion''). 

Because not all scenarios were personally relevant to all respondents, we asked respondents to determine which of two explanations a consumer would prefer based on the information provided, rather than asking them which explanation they would prefer themselves (e.g., ``If you were Vero which of the following two explanations (shown below) that explain why you are seeing this advertisement, would you prefer?''). Given the well known false consensus effect \cite{marks1987ten}, respondents' answers should be influenced by their own preferences.\footnote{Beyond the false consensus effect, there is empirical evidence in the literature for both discrepancies and similarities in how people make decisions for others vs. themselves 
	\cite{Fagerlin,Polman}. We acknowledge this limitation of our study, and the need for additional studies in which participants directly experience the AI-determined outcome rather than making choices on behalf of a third person.} 
Nevertheless, our estimate of the link between CRT and preferences may be viewed as conservative, as it really captures the link between the respondent's CRT score and their beliefs on what explanation another consumer would prefer given the vignette. 

In Study $1$, each participant answered two CBC surveys in sequence, corresponding to the two application contexts (online targeted advertising and loan applications). For targeted advertising, explanations revealed why a specific ad was shown to a user. For loan applications, explanations revealed why a loan application was rejected. In each context, each respondent provided responses to six binary choice questions.

In Study $2$, we looked into different scenarios within the loan application context. We made a distinction between positive and negative decisions (i.e., loan granted vs. denied) on the one hand, and low-stake (loan to buy a laptop, ranging from \$900 to \$2,500) and high-stake (mortgage loan ranging from \$100k to \$180k) decisions on the other hand. Combining these dimensions, each respondent was randomly assigned, between subjects, to one of the four conditions corresponding to one combination of (positive vs. negative decision) $\times$ (low- vs. high-stake). Each respondent answered six choice questions.  Distinguishing between positive and negative decisions allowed us to test our prediction that counterfactual explanations should be relatively preferred for negative decisions. Both factors (positive/negative and high/low stakes) further allow us to rule out some alternative explanations for the findings from Study 1 related to preferences for specificity. If users prefer lower-level features in explanations for loan decisions irrespective of the stakes and the positivity of the decision, then these factors are unlikely to explain differences between loan application and online advertising in Study 1.
The Web Appendix 
contains all CBC questions from both studies.

\begin{center}
	INSERT FIGURES \ref{fig:QuestionTA} and \ref{fig:QuestionCS} HERE
\end{center}

\subsection{Data Collection}
The experiments were run on the online platform \textit{Qualtrics} and consisted of the following steps. First, respondents received survey instructions, followed by a description of the explanation formats and a short quiz to test their understanding. The next step was the main task, which consisted of six choice tasks, each with two alternatives (in Study 1, each respondent completed two sets of six choice tasks). The alternatives in each task were randomly shown on the left vs. right. \textbf{Figs. \ref{fig:QuestionTA}} and \textbf{\ref{fig:QuestionCS}} show the survey interface of the main task, respectively for the online advertising and loan application contexts. After the respondents completed the main task, they were shown another attention check, completed the cognitive reflection test (CRT) and some final questions on their socio-demographics and education level. The details of the survey questions are available in the Web Appendix.

For both studies, sample sizes were set in advance, and analyses were not conducted until all data was collected. A priori, we determined the following reasons for excluding participants: (1) they did not pass the attention checks or (2) they did not complete the survey. We initially recruited $250$ respondents for Study $1$ and $1,000$ participants ($250$ per scenario) for Study $2$ on Amazon.com's Mechanical Turk (MTurk) platform. 
After screening, $216$ participants remained for the final preference analysis of Study $1$. For Study $2$, the number of respondents for the four scenarios was $244$ (high-stake/negative), $242$ (high-stake/positive), $233$ (low-stake/negative) and $241$ (low-stake/positive). \textbf{Table~\ref{table:description}} shows an overview of the survey data.

\subsection{Data Analysis}
We analyzed the data using Hierarchical Bayes (HB), the standard estimation method for CBC data~\cite{Orme2000,Allenby2005}. We modeled the prior on the part-worths as a function of the CRT score. We standardized the CRT scores, such that the main effect captures the population mean at the average CRT score and the coefficients for CRT reflect average differences in preferences related to a one standard deviation change in CRT score. This means our heterogeneity specification was as follows: 
\begin{equation}
\beta_{h} \sim N(\Delta'z_{h}, V_{\beta}),
\label{eq:2} 
\end{equation}
where $\beta_h$ represent the prior on the part-worths for respondent $h$, $z_h$ contains an intercept and the CRT score, and $\Delta$ is a matrix with two rows, that capture respectively the expected part-worths of the average respondent and the effect of a 1 standard deviation change in CRT on the expected part-worths.
We estimated the HB models in \textit{R} software using the \textit{ChoiceModelR} package~\cite{Sermas2012}. More details on the HB model specification and estimation are described in the \textbf{Appendix}.

\section{Empirical Results}
\vspace*{5mm}

\subsection{Study 1}
The estimates of the expected part-worths, i.e., point estimates of the matrix $\Delta$ in Equation \ref{eq:2},  are shown in \textbf{Table~\ref{table:results1}}.\footnote{Note that the part-worths associated with the levels of each attribute sum to zero, because we used effects coding for the attribute levels. 
} 
We construct credible intervals for relevant contrasts (i.e., comparisons between pairs of levels of the same attribute) using the last 5,000 (converged) MCMC iterations. With a slight abuse of language, we refer to one level being ``statistically significantly'' larger than another if the 95\% credible interval of the difference does not include 0.   

\subsubsection*{Format.}
For targeted advertising, one-sided importance rankings are significantly preferred over counterfactual explanations and two-sided importance rankings. While we did not have a prediction regarding one-sided vs. two-sided importance rankings, these results confirm our predictions that importance rankings should mostly be preferred over counterfactual explanations. For loan applications, counterfactual explanations are slightly preferred over two-sided importance rankings, but none of the differences are significant. Study 1 included only 
negative loan decisions (for which we predicted counterfactuals would be relatively more preferred). In Study 2, we disentangle the effects for positive (for which we predicted counterfactuals would be less preferred) vs. negative decisions.

\subsubsection*{Complexity.} In both contexts, explanations with more complexity (six features) are significantly preferred over those with less complexity (three features). This suggests that while preferences for complexity should be expected to go down as the number of features gets too large, consumers actually prefer moderate complexity over low complexity. 

\subsubsection*{Specificity.} Low-level features are preferred over high-level features in the context of loan application. 
For targeted advertising, on the other hand, participants show a preference for high-level features (e.g., ``sports websites'') over low-level features (e.g., ``basketballnews.com''). These results are consistent with our predictions 
that low-level features should be preferred in contexts that are further down the consumer's purchase process, like a loan decision (the consumer already went through the application process), which are likely construed at a lower level, and that high-level features should be preferred in contexts that are relevant at earlier stages of the purchase process, like online advertising, which should be construed at a higher level. 

\subsubsection*{CRT.} We find that people with higher CRT scores (i.e., more deliberative cognitive style) indicate a stronger preference for low-level features in the context of loan application. We find no effect of CRT on preferences for specificity in the online advertising context, or on preferences for complexity. As mentioned above, these results are conservative as respondents were asked to assess the preferences of a hypothetical consumer, which should only indirectly reflect their own preferences. 

\subsection{Study 2}
The estimated coefficients are shown in \textbf{Table~\ref{table:results_study2_positive}} (positive loan decisions) and \textbf{Table~\ref{table:results_study2_negative}} (negative loan decisions). 

\subsubsection*{Format.} We find that importance rankings are preferred over counterfactual explanations for positive decisions, but that there is no significant difference between the three formats for negative decisions. 
These results are consistent with our predictions that counterfactual explanations should be relatively more popular (or in this case, less unpopular) when the explanation is provided for a negative decision, compared to a positive decision. 


\subsubsection*{Complexity.} We again find that moderate-complexity explanations are preferred over low-complexity explanations, which supports the findings of Study $1$. Only for low-stake, negative decisions, is the difference not significant. 

\subsubsection*{Specificity.} In all scenarios, the respondents significantly prefer low-level specificity over high-level specificity, which is in line with the results in Study $1$ (for loan applications). This suggests that the differences in preferences for specificity observed in Study 1 between  online advertising and loan applications, were not due to the fact that loan application decisions were negative and that some had higher stakes. While we cannot rule out all possible alternative explanations, this further confirms that the results from Study 1 are consistent with our predictions 
that users prefer low-level features in contexts that are further down the purchase process (such as loan application decisions), and high-level features in context that occur earlier in the process (such as online display advertising).  

\subsubsection*{CRT.} Consistent with Study 1, we find that people with higher CRT scores (i.e., more deliberative cognitive style) have a stronger preference for low-level features. 

\begin{center}
	INSERT TABLES \ref{table:description} TO \ref{table:results_study2_negative} HERE
\end{center}

\section{Discussion and Future Research}

Explaining algorithmic decisions is emerging as a potential new business requirement, driven by regulations as well as consumer expectations. However, while a number of XAI algorithms have been developed, little is known about how people may react to the explanations generated by these algorithms. We treated explanations as ``products'' with  attributes being key characteristics of current XAI explanation methods (format, complexity and specificity) and studied people's preferences for these attributes using  choice-based conjoint analysis. 

Our empirical results show that importance rankings are generally preferred over counterfactual explanations, unless they are provided for negative decisions. Second, people tend to accept a moderate level of explanation complexity, irrespective of the decision context. Third, preferences regarding specificity depend on the application context (high-level for targeted advertising and low-level for loan application). In addition, the preference for low-level specificity is higher for people with a more deliberative cognitive style. Our findings are largely in agreement with predictions based on consumer behavior theories, which supports the use of Marketing theories and methods in understanding preferences for explanations. Moreover, our results provide guidance on how to design and deploy XAI algorithms while also considering the application context and the cognitive style of the end user. 

We close by highlighting a few additional potential areas for future research. First, we only studied preferences for three of the main design characteristics of XAI explanation methods and only in two application contexts. Having demonstrated the applicability of  Marketing methods in this emerging field, a natural next step would be to extend the range of both methodologies and contexts. For example, it may be possible to use feedback on explanations of, say, product recommendations to improve future recommendations or inform the design of products. Combining product recommendations with explanations and user feedback may also potentially help develop better customer insights.
Second, our results related to preferences for format and for specificity were consistent with popular theories from psychology, but we cannot rule out all alternative accounts. For example, counterfactual explanations differ from importance ranking explanations on several additional dimensions (e.g., visual, textual, double negation, presence of numerical weights). Future research might pinpoint the source of heterogeneity in preferences, which might suggest ways to improve each type of format (e.g., adding a graphical component to counterfactuals).  
We studied preferences without considering the impact of different types of explanations on outcomes such as whether people would actually accept an algorithmic decision or recommendation, the degree of algorithmic aversion, or the level of trust and understanding of the algorithms that drive the decisions. More research, possibly using field experiments or more realistic lab experiments, is needed to better understand the impact of different AI explanations on the behavior of individuals and business outcomes. Overall, we argue there is a rich area of research to which Marketing scholars can contribute, in the emerging field of XAI and the impact of algorithmic decisions and explanations on consumers' behavior and attitudes.

\appendix

\clearpage
\section*{Tables}
\begin{table}[ht!]
	\centering
	\caption{Description of the data.}
	\label{table:description}
	\scalebox{0.85}{
		\begin{tabular}{c|c}
			\textbf{Sample Size} & Study $1$: $216$ respondents\\
			& Study $2$: $960$ respondents\\
			& (respondents based in the United States\\
			& and recruited via the MTurk platform)\\
			\textbf{Attributes} & \textbf{Levels}\\
			Format & Counterfactual explanation,\\
			& One-sided importance ranking,\\
			& Two-sided importance ranking\\
			Complexity & Low (three features)\\
			& Moderate (six features)\\
			Specificity & Low-level vs. High-level\\
			\textbf{Cognitive reflection}  & Score on test for analytical reasoning\\
			& (CRT scores lie between $0$ and $3$)\\
	\end{tabular}}
\end{table}

\begin{table}[h]
	\centering
	\caption{Study $1$: Population mean part-worth estimates for each attribute as a function of CRT score.}
	\label{table:results1}
	\scalebox{0.8}{
		\begin{tabular}{lccccc}
			\multicolumn{2}{c}{\textbf{Context: Targeted advertising}}&\\
			\textbf{Attribute} & \textbf{Level} & \textbf{Intercept} & $\textbf{95\%}$ \textbf{CI} & \textbf{CRT score} & $\textbf{95\%}$ \textbf{CI}\\
			\noalign{\smallskip}\hline
			\\
			Format & Counterfactual & $-0.437$ & [$-0.619$, $-0.247$] & $-0.003$ & [$-0.267$, $0.252$]\\
			& One-sided IR & $\textbf{0.765}$* & [$0.508$, $1.044$] & $-0.127$ & [$-0.475$, $0.195$]\\
			& Two-sided IR & $-0.328$ & [$-0.621$, $-0.053$] & $0.130$ & [$-0.223$, $0.508$]\\
			\\
			Complexity & Low & $-0.324$ & [$-0.482$, $-0.165$] & $-0.078$ & [$-0.270$, $0.125$]\\
			& Moderate & $\textbf{0.324}$* & [$0.165$, $0.482$] & $0.078$ & [$-0.125$, $0.270$]\\
			\\
			Specificity & Low-level & $-0.174$ & [$-0.296$, $-0.046$] & $0.074$ & [$-0.097$, $0.253$]\\
			& High-level & $\textbf{0.174}$* & [$0.046$, $0.296$] & $-0.074$ & [$-0.253$, $0.097$]\\
			\\
			\multicolumn{2}{c}{\textbf{Context: Loan application}}&\\
			\textbf{Attribute} & \textbf{Level} & \textbf{Intercept} & $\textbf{95\%}$ \textbf{CI} & \textbf{CRT score} & $\textbf{95\%}$ \textbf{CI}\\
			\noalign{\smallskip}\hline
			\\
			Format & Counterfactual & $\textbf{0.131}$* & [$-0.056$, $0.353$] & $0.078$ & [$-0.237$, $0.376$]\\
			& One-sided IR & $-0.030$* & [$-0.326$, $0.310$]  & $-0.036$ & [$-0.354$, $0.276$]\\
			& Two-sided IR & $-0.101$* & [$-0.456$, $0.199$] & $-0.043$ & [$-0.379$, $0.271$]\\
			\\
			Complexity & Low & $-0.461$ & [$-0.639$, $-0.310$] & $0.056$ & [$-0.139$, $0.274$]\\
			& Moderate & $\textbf{0.461}$* & [$0.310$, $0.639$] & $-0.056$ & [$-0.274$, $0.139$]\\
			\\
			Specificity & Low-level & $\textbf{0.340}$* & [$0.191$, $0.487$] & $0.199^{x}$ & [$0.042$, $0.358$]\\
			& High-level & $-0.340$ & [$-0.487$, $-0.191$] & $-0.199^{x}$ & [$-0.358$, $-0.042$]\\
	\end{tabular}}
	\begin{tablenotes}
		\small
		\item The CRT scores are standardized. Credible intervals are based on the last (converged) $5000$ MCMC draws. A * indicates that the average partworth is highest among all levels of that attribute (the highest partworth is also indicated in boldface) or that the 95\% CI of the difference with the highest partworth includes $0$. A $x$ indicates that the 95\% CI of the coefficient for CRT does not include $0$. 
	\end{tablenotes}
\end{table}

\begin{table}[h]
	\centering
	\caption{Study $2$ (Positive loan decisions): Population mean part-worth estimates for each attribute as a function of CRT score.}
	\label{table:results_study2_positive}
	\scalebox{0.8}{
		\begin{tabular}{lccccc}
			\multicolumn{2}{c}{\textbf{Context: Low-stake \& positive decision ($N$=$241$)}}&\\
			\textbf{Attribute} & \textbf{Level} & \textbf{Intercept} & $\textbf{95\%}$ \textbf{CI} & \textbf{CRT score} & $\textbf{95\%}$ \textbf{CI}\\
			\noalign{\smallskip}\hline
			\\
			Format & Counterfactual & $-0.917$ & [$-1.232$, $-0.673$] & $-0.478^{x}$ & [$-0.830$, $-0.164$]\\
			& One-sided IR & $0.044$* & [$-0.442$, $0.473$] & $0.151$ & [$-0.160$, $0.463$]\\
			& Two-sided IR & $\textbf{0.873}$* & [$0.377$, $1.411$] & $0.326$ & [$-0.086$, $0.770$]\\
			\\
			Complexity & Low & $-0.383$ & [$-0.589$, $-0.162$] & $0.134$ & [$-0.079$, $0.383$]\\
			& Moderate & $\textbf{0.383}$* & [$0.162$, $0.589$] & $-0.134$ & [$-0.383$, $0.079$]\\
			\\
			Specificity & Low-level & $\textbf{0.493}$* & [$0.260$, $0.713$] & $0.175^{x}$ & [$0.005$, $0.357$]\\
			& High-level &  $-0.493$ & [$-0.713$, $-0.260$] & $-0.175^{x}$ & [$-0.357$, $-0.005$]\\
			\\
			\noalign{\smallskip}\hline
			\\
			\multicolumn{2}{c}{\textbf{Context: High-stake \& positive decision ($N$=$242$)}}&\\
			\textbf{Attribute} & \textbf{Level} & \textbf{Intercept} & $\textbf{95\%}$ \textbf{CI} & \textbf{CRT score} & $\textbf{95\%}$ \textbf{CI}\\
			\noalign{\smallskip}\hline
			\\
			Format & Counterfactual & $-0.996$ & [$-1.346$, $-0.749$] & $-0.203$ & [$-0.560$, $0.112$]\\
			& One-sided IR & $0.396$* & [$-0.020$, $0.929$] & $-0.050$ & [$-0.424$, $0.337$]\\
			& Two-sided IR & $\textbf{0.600}$* & [$0.092$, $0.965$] & $0.253$ & [$-0.173$, $0.663$]\\
			\\
			Complexity & Low & $-0.350$ & [$-0.581$, $-0.081$] & $-0.101$ & [$-0.316$, $0.159$]\\
			& Moderate & $\textbf{0.350}$* & [$0.081$, $0.581$] & $0.101$ & [$-0.159$, $0.316$]\\
			\\
			Specificity & Low-level & $\textbf{0.483}$* & [$0.289$, $0.726$] & $0.362^{x}$ & [$0.189$, $0.552$]\\
			& High-level & $-0.483$ & [$-0.726$, $-0.289$] & $-0.362^{x}$ & [$-0.552$, $-0.189$]\\
			\\
			\noalign{\smallskip}\hline
	\end{tabular}}
	\begin{tablenotes}
		\small
		\item The CRT scores are standardized. Credible intervals are based on the last (converged) $5000$ MCMC draws. A * indicates that the average partworth is highest among all levels of that attribute (the highest partworth is also indicated in boldface) or that the 95\% CI of the difference with the highest partworth includes $0$. A $x$ indicates that the 95\% CI of the coefficient for CRT does not include $0$.
	\end{tablenotes}
\end{table}

\begin{table}[h]
	\centering
	\caption{Study $2$ (Negative loan decisions): Population mean part-worth estimates for each attribute as a function of the CRT scores.}
	\label{table:results_study2_negative}
	\scalebox{0.8}{
		\begin{tabular}{lccccc}
			\multicolumn{2}{c}{\textbf{Context: Low-stake \& negative decision ($N$=$233$)}}&\\
			\textbf{Attribute} & \textbf{Level} & \textbf{Intercept} & $\textbf{95\%}$ \textbf{CI} & \textbf{CRT score} & $\textbf{95\%}$ \textbf{CI}\\
			\noalign{\smallskip}\hline
			\\
			Format & Counterfactual & $-0.154$* & [$-0.377$, $0.034$] & $0.009$ & [$-0.232$, $0.244$]\\
			& One-sided IR & $0.038$* & [$-0.246$, $0.294$] & $-0.146$ & [$-0.454$, $0.150$]\\
			& Two-sided IR & $\textbf{0.116}$* & [$-0.165$, $0.432$] & $0.138$ & [$-0.153$, $0.426$]\\
			\\
			Complexity & Low & $-0.034$* & [$-0.184$, $0.130$] & $0.052$ & [$-0.113$, $0.227$]\\
			& Moderate & $\textbf{0.034}$* & [$-0.130, 0.184$] & $-0.052$ & [$-0.227$, $0.113$]\\
			\\
			Specificity & Low-level & $\textbf{0.499}$* & [$0.375$, $0.654$] & $0.309^{x}$ & [$0.151$, $0.487$]\\
			& High-level & $-0.499$ & [$-0.654$, $-0.375$] & $-0.309^{x}$ & [$-0.487$, $-0.151$]\\
			\\
			\noalign{\smallskip}\hline
			\\
			\multicolumn{2}{c}{\textbf{Context: High-stake \& negative decision ($N$=$244$)}}&\\
			\textbf{Attribute} & \textbf{Level} & \textbf{Intercept} & $\textbf{95\%}$ \textbf{CI} & \textbf{CRT score} & $\textbf{95\%}$ \textbf{CI}\\
			\noalign{\smallskip}\hline
			\\
			Format & Counterfactual & $\textbf{0.023}$* & [$-0.194$, $0.215$] & $0.029$ & [$-0.232$, $0.271$]\\
			& One-sided IR & $-0.023$* & [$-0.406$, $0.324$] & $-0.073$ & [$-0.393$, $0.279$]\\
			& Two-sided IR & $-0.001$* & [$-0.386$, $0.421$] & $0.043$ & [$-0.318$, $0.387$]\\
			\\
			Complexity & Low & $-0.283$ & [$-0.441$, $-0.116$] & $0.047$ & [$-0.132$, $0.235$]\\
			& Moderate & $\textbf{0.283}$* & [$0.116$, $0.441$] & $-0.047$ & [$-0.235$, $0.132$]\\
			\\
			Specificity & Low-level & $\textbf{0.661}$* & [$0.517$, $0.821$] & $0.383^{x}$ & [$0.205$, $0.575$]\\
			& High-level & $-0.661$ & [$-0.821$, $-0.517$] & $-0.383^{x}$ & [$-0.575$, $-0.205$]\\
			\\
			\noalign{\smallskip}\hline
	\end{tabular}}
	\begin{tablenotes}
		\small
		\item The CRT scores are standardized. Credible intervals are based on the last (converged) $5000$ MCMC draws. A * indicates that the average partworth is highest among all levels of that attribute (the highest partworth is also indicated in boldface) or that the 95\% CI of the difference with the highest partworth includes $0$. A $x$ indicates that the 95\% CI of the coefficient for CRT does not include $0$.
	\end{tablenotes}
\end{table}

\clearpage
\section*{Figures}

\clearpage
\section*{Appendix: Hierarchical Bayes Model Specification and Estimation}


\label{Appendix:HB}

The attributes and attribute levels under investigation are shown in \textbf{Table \ref{table:description}}. In Study 1 (resp., Study 2) each respondent was asked $12$ (resp., $6$) paired comparisons. In all, after cleaning the survey data, for Study 1 (resp., Study 2) we had a total of $2,592$ (resp., $5,760$) paired comparisons from $216$ (resp., $960$) respondents available for the analysis. The CRT scores of each respondent were also available. The choice probabilities for the binary logit model can be expressed as:

\begin{equation}
Pr(y_{hi}=1) = \frac{exp[(x_{hi1}-x_{hi2})'\beta_{h}]}{1 + exp[(x_{hi1}-x_{hi2})'\beta_{h}]}
\label{eq:1} 
\end{equation}

where $h$ denotes the respondent, $i$ denotes the paired comparison, $Pr(y_{hi}=1)$ denotes the probability of selecting the first explanation, $Pr(y_{hi}=0)$ is the probability of selecting the second explanation, $x_{hi1}$ and $x_{hi2}$ are attribute vectors of dimension $4$ with elements taking on values to denote the absence or presence of an attribute level\footnote{There are $4$ columns in $x_{hi}$ that contain the differences in the attribute vectors ($x_{hi1}$ and $x_{hi2}$).}, and $\beta_{h}$ represents a vector of part-worth utilities to be estimated. The choice data contains a respondent identifier ($h$), a choice indicator ($y$) and the levels of each attribute for the explanations in the comparison. The choice indicator takes on values of $1$ or $0$, with a $1$ indicating that the first explanation was selected by the respondent. Both the (de-identified) choice data and CRT scores are available in the Web Appendix on Github\footnote{
	url currently undisclosed, for blinded review.
}.





We included the three attributes with their levels in the model and treated them as categorical (effects coding). The prior distributions of the model coefficients are assumed to be normally distributed. We had no prior expectations for the (mean) utility values for the attribute levels, which is equivalent to using prior mean values of zero. The prior variance of the HB model was set to $2$ and the prior degrees of freedom to $5$ (default settings of the \textit{ChoiceModelR} package in $R$). These settings indicate that much information for the individual-level part-worth estimates is borrowed from the population parameters. 
The priors of the means of the distributions of $\beta$ and $\Delta$ are respectively a vector of zeros with length $4$, and $0$. We set the number of draws from the Markov chain Monte Carlo (MCMC) to $20,000$ and keep the last $5,000$ draws for the analysis \cite{Allenby2005}.

\clearpage

\end{document}